\documentclass[11pt,twoside,a4paper,cmspaper,final,collab]{cms-tdr}
\def\svnVersion{a4e0102-D}\def\svnDate{2026/08/10}\def\cmsCernNoTag{CERN-EP-2025-163}\def\cmsCernDate{\today}\def\cmsMessage{Published in Physical Review Letters as \href{http://dx.doi.org/10.1103/hgvs-jplv}{\doi{10.1103/hgvs-jplv}.}}
\begin{document}\cmsNoteHeader{BPH-23-007}

\newcommand{\abseta}{\ensuremath{\abs{\eta}}\xspace}
\newcommand{\deltaphi}{\ensuremath{\Delta\phi}\xspace}
\newcommand{\deltay}{\ensuremath{\Delta y}\xspace}
\newcommand{\deltaR}{\ensuremath{\Delta R}\xspace}
\newcommand{\StextIP}{\ensuremath{S_\text{IP}}\xspace}
\newcommand{\upsi}{\PGU}
\newcommand{\upsis}{\ensuremath{\PGUP{1\mathrm{S}}}\xspace}
\newcommand{\upsins}{\ensuremath{\PGUP{\mathit{n}\mathrm{S}}}\xspace}
\newcommand{\pp}{\ensuremath{\Pp\Pp}\xspace}
\newcommand{\Zupsis}{\ensuremath{\PZ+\upsis}\xspace}
\newcommand{\Zupsns}{\ensuremath{\PZ+\upsins}\xspace}
\newcommand{\mumu}{\ensuremath{\PGmp\PGmm}\xspace}
\newcommand{\mumumumu}{\ensuremath{\mumu\mumu}\xspace}
\newcommand{\Zmumumumu}{\ensuremath{\PZ\to\mumu\mumu}\xspace}
\newcommand{\tomumu}{\ensuremath{\to\mumu}\xspace}
\newcommand{\fourmu}{\ensuremath{4\PGm}\xspace}
\newcommand{\mfourmu}{\ensuremath{m_{\fourmu}}\xspace}
\newcommand{\mtwomu}{\ensuremath{m_{\mumu}}\xspace}
\newcommand{\ZToFourMu}{\ensuremath{\PZ\to\fourmu}\xspace}
\newcommand{\Zmumu}{\ensuremath{\PZ\to\mumu}\xspace}
\newcommand{\UpsiTwoS}{\ensuremath{\PGUP{2\mathrm{S}}}\xspace}
\newcommand{\ZUpsiTwoS}{\ensuremath{\PZ+\UpsiTwoS}\xspace}
\newcommand{\UpsiThreeS}{\ensuremath{\PGUP{3\mathrm{S}}}\xspace}
\newcommand{\ZUpsiThreeS}{\ensuremath{\PZ+\UpsiThreeS}\xspace}
\newcommand{\Upsimumu}{\ensuremath{\upsis\to\mumu}\xspace}
\newcommand{\isolation}{\ensuremath{\mathcal{I}}\xspace}
\newcommand{\RUpsi}{\ensuremath{\mathcal{R}_{\Zupsis}}\xspace}
\newcommand{\RUpsiDPS}{\ensuremath{\mathcal{R}^\text{DPS}_{\Zupsis}}\xspace}
\newcommand{\BZmumumumu}{\ensuremath{\mathcal{B}(\ZToFourMu)}\xspace}
\newcommand{\BZmumu}{\ensuremath{\mathcal{B}(\Zmumu)}\xspace}
\newcommand{\BUpsimumu}{\ensuremath{\mathcal{B}(\Upsimumu)}\xspace}
\newcommand{\sigmappZupsis}{\ensuremath{\sigma(\pp\to\PZ+\upsis)}\xspace}
\newcommand{\sigmappZupsisBRs}{\ensuremath{\sigmappZupsis \BZmumu \BUpsimumu}\xspace}
\newcommand{\sigmappZ}{\ensuremath{\sigma(\pp\to\PZ)}\xspace}
\newcommand{\sigmappZBRshort}{\ensuremath{\sigmappZ\BZmumumumu}\xspace}
\newcommand{\mtwomuPZ}{\ensuremath{\mtwomu^\PZ}\xspace}
\newcommand{\mtwomuupsi}{\ensuremath{\mtwomu^{\upsis}}\xspace}
\newcommand{\AZupsis}{\ensuremath{A_{\Zupsis}}\xspace}
\newcommand{\AZ}{\ensuremath{A_{\PZ}}\xspace}

\newcommand{\crosssectionformula}{\ensuremath{\left(\frac{N^\text{SPS}_{\Zupsis}}{\epsilon^\text{SPS}_{\Zupsis}}
                                    + \frac{N^\text{DPS}_{\Zupsis}}{\epsilon^\text{DPS}_{\Zupsis}} \right)
                                    \frac{\epsilon_{\PZ}}{N_{\PZ}}
                                             }\xspace}

\newcommand{\RUpsiresult}{\ensuremath{\mathcal{R}_{\Zupsis}=(21.1 \pm 5.5 \stat\pm 0.6 \syst)\ten{-3}}\xspace}
\newcommand{\RUpsiresultDPS}{\ensuremath{\mathcal{R}_{\Zupsis}^{\text{DPS}}=(20.2 \pm 5.4 \stat\pm 0.6 \syst)\ten{-3}}\xspace}

\newcommand{\sigmaeff}{\ensuremath{\sigma_\text{eff}}\xspace}
\newcommand{\sqrts}{\ensuremath{\sqrt{s}}\xspace}
\newcommand{\yield}{\ensuremath{34.6\pm 9.0}\xspace}
\newcommand{\DPSFraction}{\ensuremath{f_\text{DPS}=0.95 \pm 0.25}\xspace}
\newcommand{\sigmaeffresult}{\ensuremath{13.0^{+7.7}_{-3.4}\unit{mb}}\xspace}

\newcommand{\spsyield}{\ensuremath{N^\text{SPS}_{\Zupsis}=1.7 \pm 8.7}\xspace}
\newcommand{\dpsyield}{\ensuremath{N^\text{DPS}_{\Zupsis}=32.9 \pm 8.7}\xspace}

\newcommand{\UpsiSigModUnc}{1.8\%\xspace}
\newcommand{\UpsiBkgModUnc}{0.3\%\xspace}
\newcommand{\ZSigModUncZUpsi}{1.1\%\xspace}
\newcommand{\ZSigModUncZToMuMuMuMu}{1.6\%\xspace}
\newcommand{\ZBkgModUncZUpsi}{0.7\%\xspace}
\newcommand{\ZBkgModUncZToMuMuMuMu}{0.2\%\xspace}
\newcommand{\MCStatsUncRatio}{0.2\%\xspace}
\newcommand{\MCStatsUncSigmaEff}{0.2\%\xspace}
\newcommand{\BFUpsiMuMuUnc}{1.6\%\xspace}
\newcommand{\BFZMuMuUnc}{0.2\%\xspace}
\newcommand{\BFZMuMuMuMuUnc}{4.0\%\xspace}
\newcommand{\UpsiXSecTheoryUnc}{1.9\%\xspace}
\newcommand{\RatioTotalSystUncert}{2.7\%\xspace}
\newcommand{\SigmaEffTotalSystUncert}{5.5\%\xspace}

\newcommand{\PGcbJ}[1]{{\HepParticleResonanceFull{\PGc}{\PQb \mathit{J}}{}{#1}{}{}}\Xspace}

\newlength{\cmsFigSize}
\ifthenelse{\boolean{cms@external}}{
    \providecommand{\cmsLeft}{top\xspace}
    \providecommand{\cmsRight}{bottom\xspace}
    \setlength{\cmsFigSize}{0.225\textwidth}

    \renewcommand{\appendixname}{End Matter}
}{
    \providecommand{\cmsLeft}{left\xspace}
    \providecommand{\cmsRight}{right\xspace}
    \setlength{\cmsFigSize}{0.48\textwidth}
}
\providecommand{\suppMaterial}{\appendixname~\ref{app:suppMat}}
\newlength\cmsTabSkip\setlength{\cmsTabSkip}{1ex}
\providecommand{\cmsTable}[1]{\resizebox{\textwidth}{!}{#1}}

\cmsNoteHeader{BPH-23-007}
\title{Observation of \texorpdfstring{\upsis}{Y(1S)} + \texorpdfstring{\PZ}{Z} associated production and measurement of the effective double-parton scattering cross section in proton-proton collisions at \texorpdfstring{$\sqrts=13\TeV$}{sqrt(s) = 13 TeV}}
\date{\today}

\abstract{
The observation of associated production of an \upsis meson with a \PZ boson and a measurement of the ratio of its fiducial cross section to the fiducial cross section of the \PZ boson are presented. Both the \upsis meson and the \PZ boson are identified via decays into a pair of opposite-sign muons. The analysis is based on proton-proton (\pp) collision data at $\sqrt{s}=13\TeV$, collected with the CMS detector in 2016--2018 and corresponding to an integrated luminosity of 138\fbinv. Using the production of the \PZ boson decaying into four muons as a normalization channel, the ratio of the fiducial cross sections \sigmappZupsisBRs to \sigmappZBRshort is measured to be \RUpsiresult, where stat and syst denote the statistical and systematic uncertainties, respectively. The result is used to extract the effective double-parton scattering cross section $\sigmaeff = \sigmaeffresult$. In addition, for the first time, \sigmaeff is measured in bins of the transverse momentum of the \upsis meson or of the \PZ boson. 
}

\hypersetup{
pdfauthor={CMS Collaboration},
pdftitle={Observation of Y(1S) + Z associated production and measurement of the effective double-parton scattering cross section in proton-proton collisions at sqrt(s) = 13 TeV},
pdfsubject={CMS},
pdfkeywords={CMS, b physics, standard model, proton structure, DPS, SPS}}

\maketitle 

Despite extensive studies, the exact production mechanism of heavy quarkonia in hadron collisions has yet to be fully understood~\cite{Andronic:2015wma, Brambilla:2010cs}. From a theoretical standpoint, quarkonium production is a two-step process. The first step involves the creation of a heavy quark-antiquark pair, while the second step involves its hadronization into a physical quarkonium state~\cite{Lansberg:2019adr}. One can describe the first step using the methods of perturbative quantum chromodynamics, while the second step is inherently nonperturbative and therefore presents a much greater theoretical challenge. This is mainly because of difficulties in the calculation of long-distance matrix elements that express the hadronization probability of a heavy quark-antiquark pair into a quarkonium state~\cite{Bodwin:1994jh}. 

The picture becomes even more complex for a quarkonium state produced in association with an electroweak boson, where none of the theoretical models can capture all aspects of the experimental results~\cite{ATLAS:2019jzd, ATLAS:2014ofp,ATLAS:2014yjd,Acosta:2003mu,Gong:2012ah,PhysRevD.83.014001,Song:2011pxf}. A major complication in the description of these processes stems from the interplay of two different production mechanisms. A quarkonium state in association with an electroweak boson can be produced either through the interaction of a single pair of partons, in a process known as single-parton scattering~(SPS), or through the interactions of two pairs of partons, known as double-parton scattering~(DPS)~\cite{Rinaldi:2018bsf,Rinaldi:2018slz,Lansberg:2016rcx,Rinaldi:2013vpa,Gong:2012ah}. The contribution of the DPS process can be quantified via a parameter known as the effective DPS cross section, \sigmaeff, which is closely related to the distribution of the partons within the nucleon in the plane transverse to the proton momentum~\cite{PhysRevD.83.114047}. 

So far, experimental measurements~\cite{CMS:2021qsn,CMS:2024wgu,CMS:2019jcb,Aaij:2016bqq,Aaboud:2016fzt,Abazov:2015fbl,Abazov:2014qba,ATLAS:2014ofp,ATLAS:2013aph,D0:2009apj,CDF:1997yfa,CDF:1993sbj,UA2:1991apc} and recent phenomenological studies~\cite{Lansberg:2019adr,Lansberg:2017chq,Lansberg:2016muq,Shao:2016wor} have provided only an integrated measurement of \sigmaeff over either a fiducial or the full phase space. The difference between the measurements of \sigmaeff for various processes is likely the result of the purely geometrical definition of \sigmaeff, which neglects parton correlations~\cite{Bansal:2014paa}. 

In this Letter, the associated production of an \upsins meson ($n = 1,2,3$) with a \PZ boson is studied, with a focus on the dominant \Zupsis production. By measuring the overall fiducial cross section for the \Zupsis process, as well as by splitting it into the SPS and DPS components, we obtain \sigmaeff both integrated over the analyzed phase space and 
in bins of \upsis meson and \PZ boson transverse momenta~(\pt). The measurement is based on proton-proton~(\pp) collision data at a center-of-mass energy of 13\TeV, collected by the CMS experiment at the CERN LHC in 2016--2018 and corresponding to an integrated luminosity of 138\fbinv~\cite{CMS-LUM-17-003,CMS-PAS-LUM-17-004,CMS-PAS-LUM-18-002}. Both the \PZ boson and the \upsins meson are reconstructed using dimuon~(\mumu) decays. In order to reduce the systematic uncertainties of the measurement, we normalize the signal cross section to that of \PZ boson production with a subsequent decay into a pair of dimuons (in what follows denoted as \fourmu)~\cite{CMS:2026xzs}, which yields an identical final state with similar kinematical properties. The details of the CMS detector, the simulated signal and background samples, and the triggers used to collect the data are given in \suppMaterial. Tabulated results are provided in the HEPData record for this analysis~\cite{hepdata}.

In the \Zupsns channel, the opposite-sign~(OS) muon pairs forming the \upsins meson and \PZ boson candidates are each fitted~\cite{Prokofiev:2005zz} to a common vertex, with the vertex probability required to exceed 5\%. 
Each muon forming the \upsins candidate must pass the \emph{soft} identification requirement~\cite{CMS:2018rym} (optimized for low-\pt muons typical of quarkonium decays) and have $\pt>3\GeV$ and pseudorapidity $\abseta < 2.4$. 
The invariant mass of the \upsins candidate is restricted to be within $8.5<\mtwomu<11\GeV$ and its rapidity is required to fall within $\abs{y}<2.4$. The two OS muons that form the \PZ candidate are required to pass the \emph{tight} identification~\cite{CMS:2018rym} (optimized for high-\pt muons), be within $\abseta < 2.4$, and have $\pt>30$ and $>$15\GeV for the leading and subleading muon, respectively. To ensure that all muons originate from the same \pp collision, the three-dimensional distance between the primary vertex and the muon track at its point of closest approach divided by its uncertainty, \StextIP, is required to be less than 4 for each muon. The primary vertex corresponds to the hardest scattering in the event, evaluated using tracking information alone, as described in Section 9.4.1 of Ref.~\cite{CMS-TDR-15-02}. The invariant mass of the \PZ candidate must lie within 66--116\GeV, where a relatively wide window was used to better constrain the background. In case both possible OS muon pairings satisfy the above requirements, which happens in less than 1\% of the events, we choose the pairing with the invariant mass of the \PZ candidate closest to the nominal \PZ boson mass~\cite{PDG}. 

All muons are required to be isolated from hadronic activity in the event. To satisfy this requirement, the scalar sum of transverse energy deposits in the calorimeters and the \pt of tracks is computed in a cone of radius $\deltaR\equiv\sqrt{\smash[b]{(\Delta\eta)^2+(\deltaphi)^2}}=0.3$ centered on the muon candidate, where \deltaphi is the difference in azimuthal angle. The \pt values of other muons from the \PZ or \upsins candidates that fall within the isolation cone are subtracted from the isolation sum. The average hadronic activity in the event, which includes the effect of multiple interactions within the same or adjacent bunch crossings~(pileup), is also subtracted from this sum~\cite{CMS:2018rym}. The ratio of the corrected sum to the muon \pt, a quantity hereafter referred to as \isolation, is required to be smaller than 0.7 or 0.35 for muons forming the \upsins or \PZ candidate, respectively. In addition, all four muons are required to originate from a common vertex with vertex fit probability $>$1\%. This final requirement plays a crucial role in the signal channel, as it effectively suppresses events where the \upsins meson and the \PZ boson originate from two independent \pp interactions, which can occur in high-pileup conditions. Due to the relatively large \upsins production cross section, the probability of such events is nonnegligible; imposing a common vertex requirement reduces this background to a negligible level.

In the \ZToFourMu normalization channel, the selection is similar to that of Ref.~\cite{CMS:2026xzs}. We require events with at least four muons and examine all the electrically neutral combinations of \fourmu that can be formed. The invariant mass of the \fourmu, \mfourmu, must lie between 80 and 100\GeV. The leading (subleading) muon must have $\pt > 20$ (10)\GeV and the two trailing muons are required to have $\pt > 5\GeV$. All four muons must be within $\abseta < 2.4$, pass the \emph{tight} identification~\cite{CMS:2018rym}, and be separated from each other by $\deltaR > 0.02$. All four muons must have \isolation less than 0.35 and $\StextIP <4$. All possible OS muon pairs in this channel must have an invariant mass $\mtwomu>1\GeV$. The four muons are combined into two pairs of OS muons that maximize the mass difference between the pairs, with each pair required to have a vertex fit probability $>$5\%. Unlike in the signal channel, we do not perform a common vertex fit for the four muons, as here the contribution from events with two muons stemming from one vertex and two muons stemming from another that has \mfourmu in the 80--100 GeV window is negligible. While multiple four-muon candidates per event are allowed, all the selected events have only a single four-muon combination that satisfies the requirements.

In the signal channel, we implement an extended unbinned maximum likelihood fit in two dimensions, to the \PZ and the \upsins candidate invariant masses. We model the \PZ resonance mass shape with a Voigtian function, a convolution of Gaussian and Breit--Wigner distributions, assuming no $\PZ/\PGg^{\ast}$ interference. The width of the Breit--Wigner function is fixed to the world-average value for the \PZ boson, 2.50\GeV~\cite{PDG}, while that of the Gaussian function is fixed to a value determined from simulation, because of the limited event count in data. The mean of the Gaussian is also fixed to the world-average value of $m_\PZ = 91.188\GeV$~\cite{PDG}. As the natural widths of the \upsins states are very small, we parameterize their mass shapes with Gaussian distributions, with widths fixed to values determined from simulation studies. The combinatorial backgrounds for each mass combination are described by exponential functions with the parameters free in the fit. We create a two-dimensional~(2D) template from these one-dimensional functions as a direct product of the functions that describe the mass spectra of the \upsins and \PZ candidates. This fit includes eight yields: three \upsins signal plus \PZ signal, three \upsins signal plus \PZ background, one \upsins background plus \PZ signal, and one \upsins background plus \PZ background.
\begin{figure}[tb!]
\centering
    \includegraphics[width=\cmsFigSize]{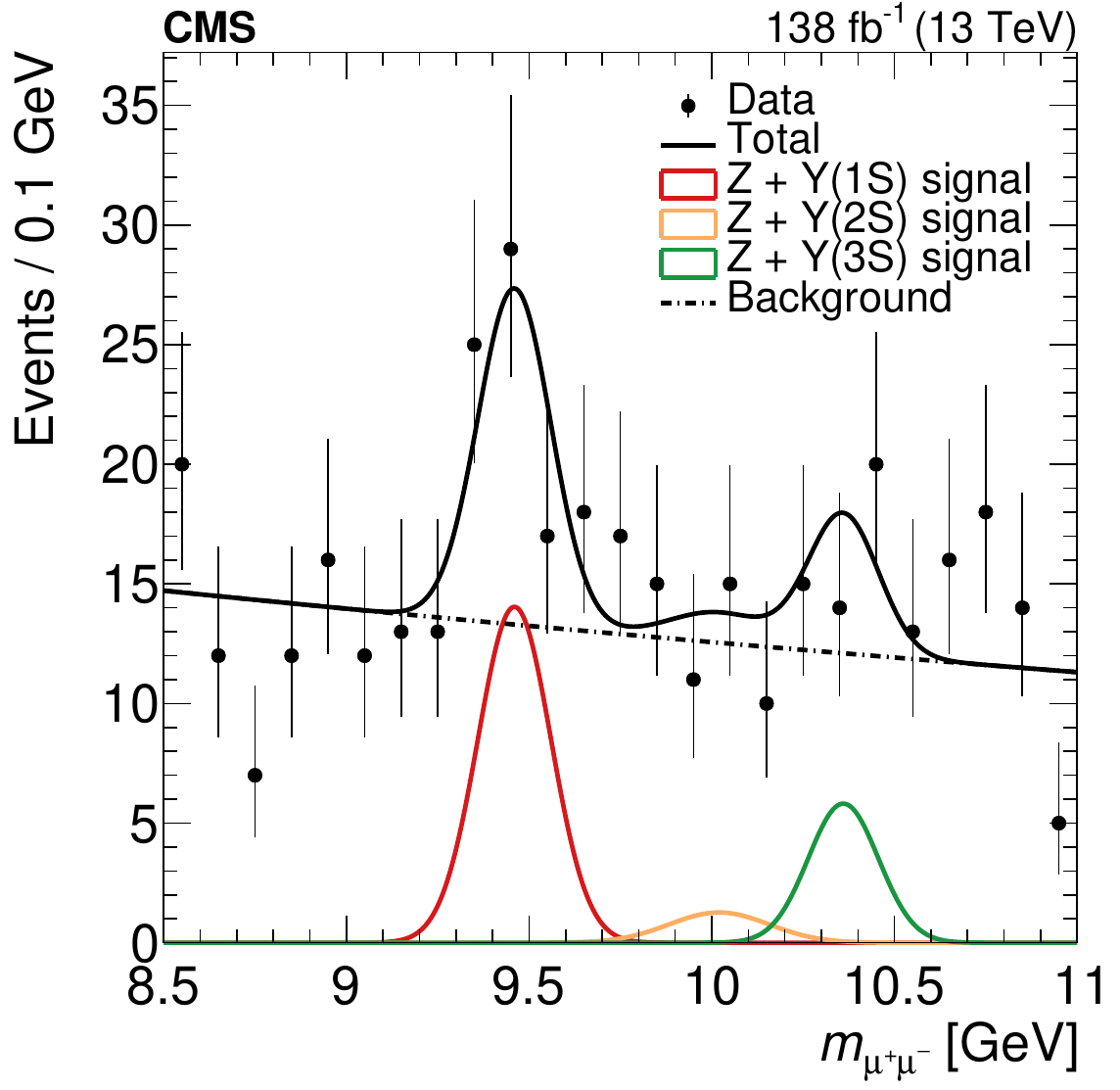}
    \includegraphics[width=\cmsFigSize]{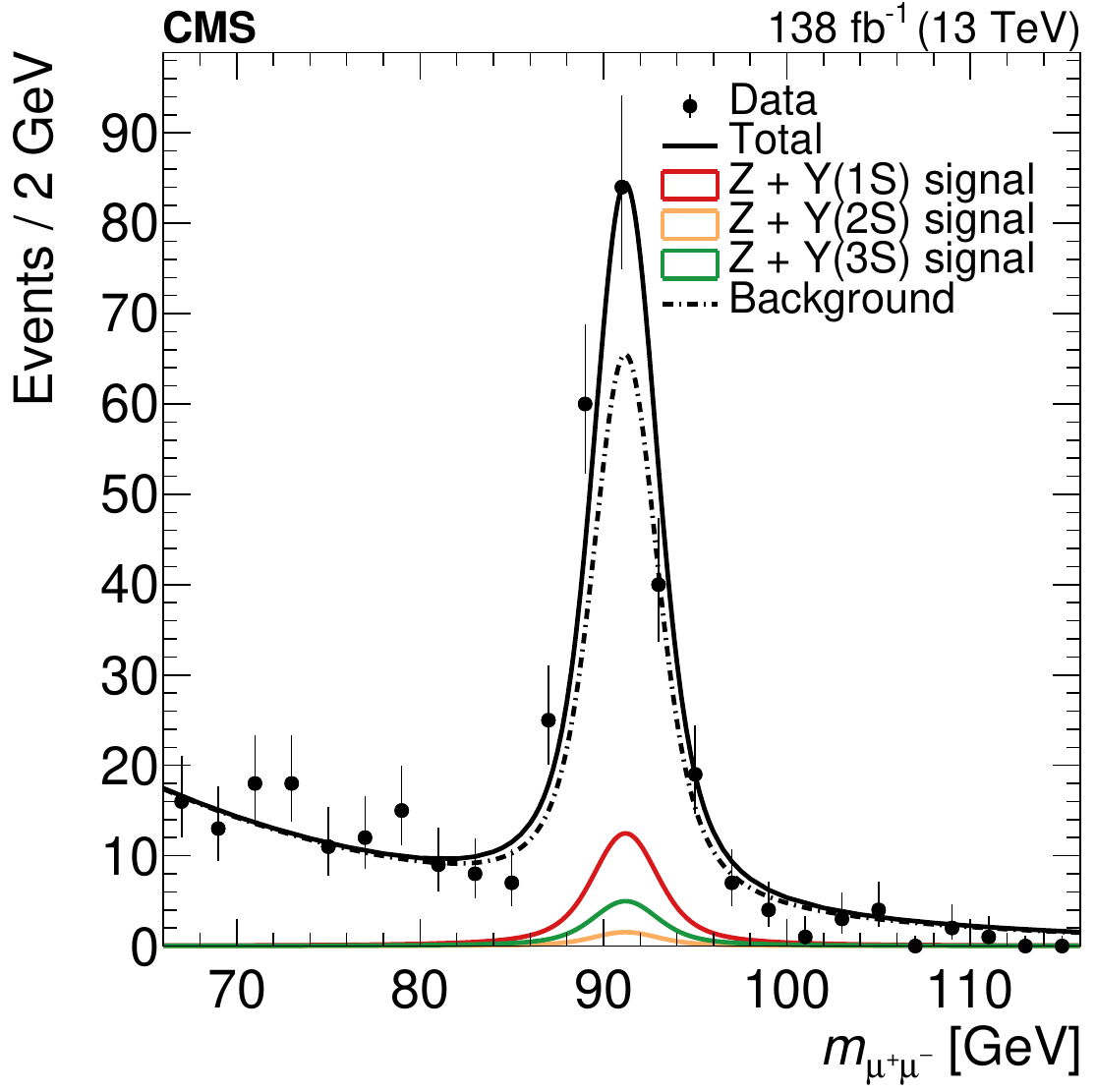}
\ifthenelse{\boolean{cms@external}}{\vspace*{-0.1in}}{}
    \caption{Dimuon invariant mass distributions for the \upsins (left) and  \PZ (right) candidates. In each panel, the data are represented by points, with vertical bars indicating the statistical uncertainties. The solid black curve is the overall fit to data. The red, yellow, and green curves correspond to the signal yields for the three \upsins states, while the dash-dotted curve represents the background. The peaking background in the right plot originates from events in which the \PZ boson candidate is genuine, while the \upsins meson candidate originates from the significant combinatorial background seen in the left plot.}
\ifthenelse{\boolean{cms@external}}{\vspace*{-0.2in}}{}
    \label{fig:ZUpsiUpsiFit}
\end{figure}

The projections of the 2D model onto the two variables of interest are shown in Fig.~\ref{fig:ZUpsiUpsiFit}. The signal yields returned from the fit are $34.6 \pm 9.0$ for \Zupsis, $2.9 \pm 6.7$ for \ZUpsiTwoS, and $13.9 \pm 7.2$ for \ZUpsiThreeS. These yields include both direct production and feed-down from heavier bottomonium states. The statistical significance of the \Zupsis signal is 5.3 standard deviations, as evaluated using Wilks' theorem~\cite{10.1214/aoms/1177732360} from the likelihood ratio of the default signal-plus-background fit and the background-only fit, by imposing $N(\Zupsis) = 0$, while treating possible contributions from \ZUpsiTwoS and \ZUpsiThreeS as part of the background. To validate the fit, we performed 500 pseudo-experiments and checked that the distribution of the pulls is well described by a Gaussian function with a zero mean and a variance of unity. The significances of the \UpsiTwoS and \UpsiThreeS resonances are below 3 standard deviations. Consequently, in what follows, we focus on the \Zupsis case.

\begin{figure}[tb!]
    \centering
    \includegraphics[width=\cmsFigSize]{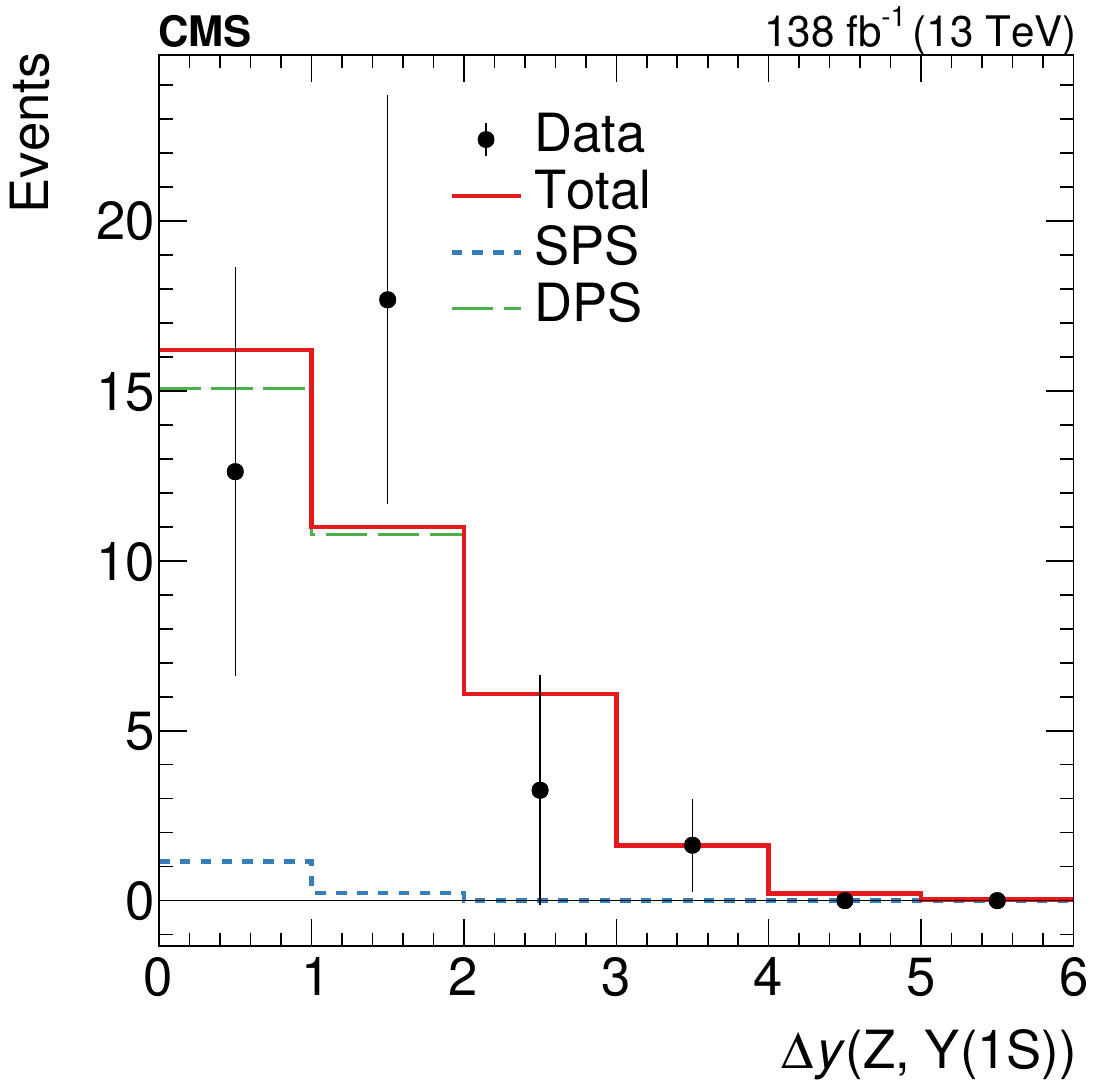}
    \includegraphics[width=\cmsFigSize]{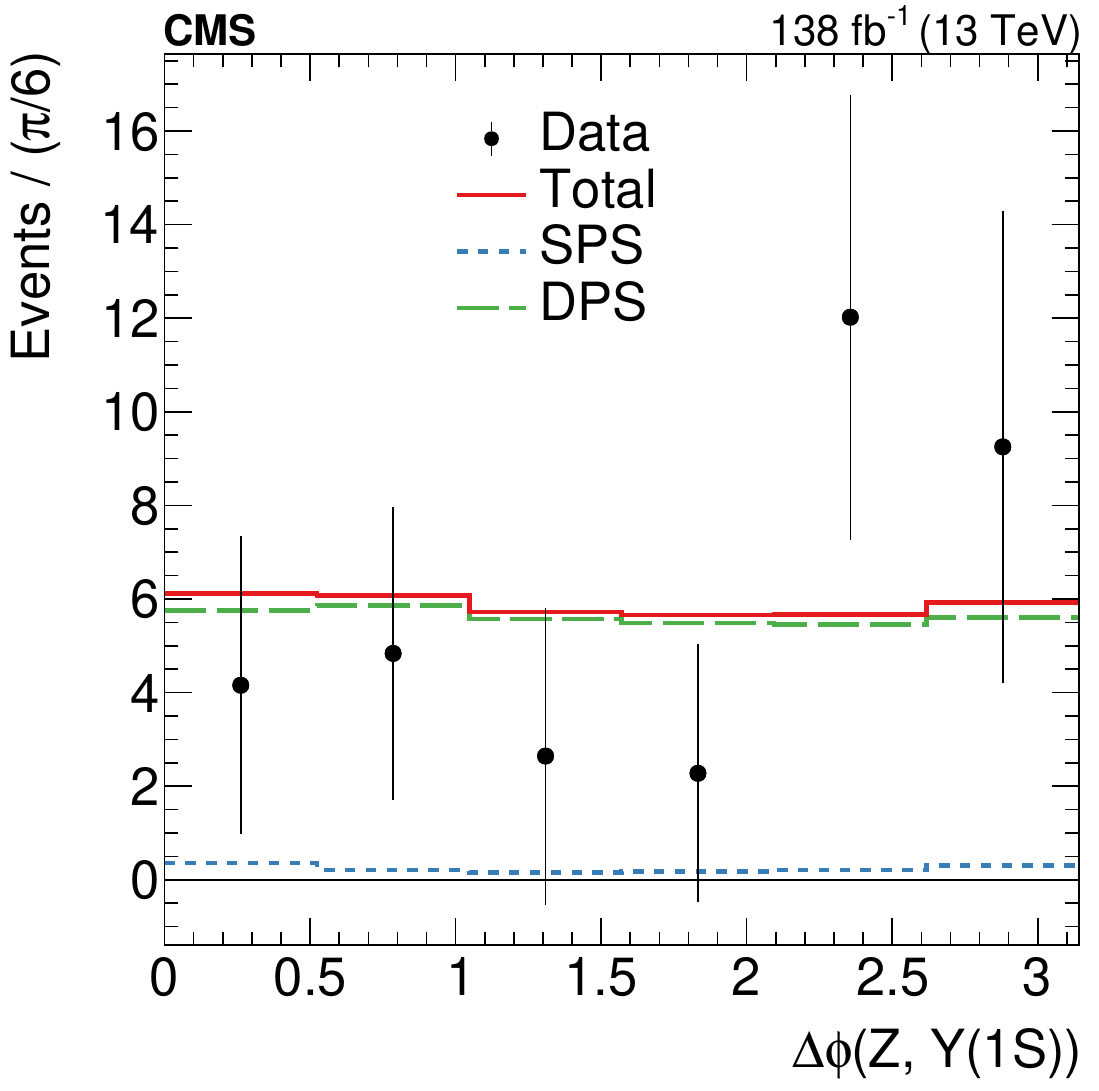}
\ifthenelse{\boolean{cms@external}}{\vspace*{-0.1in}}{}
    \caption{Distributions of the \deltay (left) and \deltaphi (right) separations between the \upsis meson and the \PZ boson in data (black points with vertical bars indicating the statistical uncertainties), and the extracted DPS (green histogram), SPS (blue histogram), and SPS+DPS (red histogram) contributions from the template fit to data. The distributions and corresponding uncertainties are obtained after the application of \textit{sPlot} weights from an unbinned template fit to the \deltay and \deltaphi variables.}
\ifthenelse{\boolean{cms@external}}{\vspace*{-0.2in}}{}
    \label{fig:deltayphi}
\end{figure}

Using the results of this fit, the \textit{sPlot} technique~\cite{Pivk:2004ty} is employed to assign weights to each event for each of the eight signal and background categories. Only the \Zupsis weights are considered signal and are used to obtain the background-subtracted \Zupsis signal distributions of the separations in rapidity (\deltay) and azimuth between the \PZ and \upsis candidates. It was confirmed that the variables used to extract the weights, namely the dimuon masses associated with the \PZ boson and \upsins candidates, and the \deltay and \deltaphi variables to which the weights are applied, are independent.  The SPS and DPS contributions are determined from an unbinned maximum likelihood fit to the 2D $(\deltay, \deltaphi)$ distribution, using 2D SPS and DPS templates obtained from simulation studies based on the kernel density estimation method~\cite{Cranmer:2000du}. The DPS template was additionally verified in data using the event-mixing method. The result of the fit is the DPS fraction, which is found to be \DPSFraction, where the uncertainty on the fraction is determined through a bootstrapping procedure. Applying the result for the DPS fraction $f_\text{DPS}$ to the total yield $N_{\Zupsis}^\text{total} =$ \yield, we find SPS and DPS yields of \spsyield and \dpsyield, respectively. 
The \deltay and \deltaphi signal distributions and the fit projections are shown in Fig.~\ref{fig:deltayphi}.

\begin{figure}[tb!]
  \centering
  \includegraphics[width=\cmsFigSize]{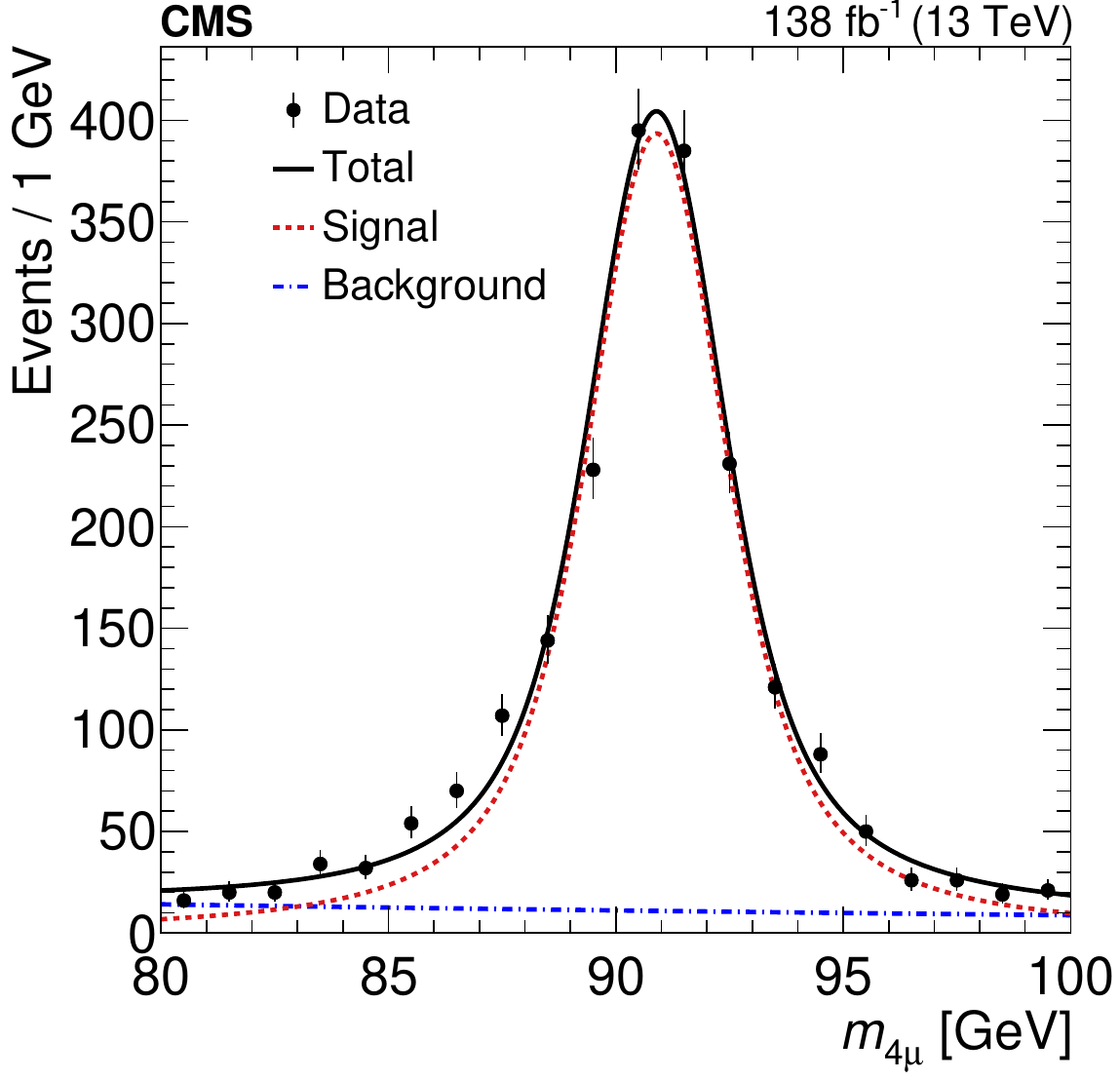}
\ifthenelse{\boolean{cms@external}}{\vspace*{-0.1in}}{}
  \caption{The \mfourmu distribution in the normalization channel with vertical bars on the data points indicating the statistical uncertainties. The result of the full fit is shown with the solid black line, while the signal and background components are shown with the dashed red and dash-dotted blue lines, respectively.}
  \ifthenelse{\boolean{cms@external}}{\vspace*{-0.2in}}{}
  \label{fig:ZOnlySigYield}
\end{figure}

In the normalization channel, we perform an extended unbinned maximum likelihood fit to the \mfourmu distribution. The \PZ boson signal is modeled with a Voigtian function, with the width of the Breit--Wigner distribution fixed to the world-average value of the \PZ boson width, and the Gaussian function parameters, including the mean, floating in the fit. The background is modeled via an exponential function with parameters floating in the fit. The resulting \PZ signal yield is $1860 \pm 60$ events. The \mfourmu distribution and the fit projection are shown in Fig.~\ref{fig:ZOnlySigYield}.

We evaluate various systematic uncertainties that affect the results by changing the model and repeating the fits. The sources of systematic uncertainties considered include the signal and background shape parameterizations of the dimuon invariant mass spectra in the signal channel and the \mfourmu spectrum in the normalization channel, limited event counts in simulated samples, uncertainties in world-average branching fractions, and uncertainty in the \upsis production cross section calculation, which only affects the \sigmaeff measurement. These uncertainties are detailed in \suppMaterial. The results are dominated by the statistical uncertainty.
\begin{table*}[tb!]
\centering
\topcaption{Summary of selection criteria applied to the signal $\PZ(\tomumu) + \upsis(\tomumu)$ and normalization \ZToFourMu channels. The ``\NA'' sign means that the selection is not applied.}
\ifthenelse{\boolean{cms@external}}{
\cmsTable{
\begin{scotch}{lcc}
Selection                           & \Zmumu + \Upsimumu                                    & \ZToFourMu     \\ \hline
All muons                           & $\abseta < 2.4$,\; $\StextIP<4$                       &  $\abseta < 2.4$,\; $\StextIP<4$     \\
Muon \pt [\GeVns{}]                 & $>$ 30, 15, 3, 3                                      & $>$ 20, 10, 5, 5   \\
Muon ID                             & \emph{tight}, \emph{tight}, \emph{soft}, \emph{soft}  & \emph{tight} for all muons \\
Isolation \isolation                & $<$ 0.35, 0.35, 0.7, 0.7                              & $<$0.35 for all muons \\
Mass ranges [\GeVns{}]              & $66 < \mtwomuPZ < 116$, $8.5 < \mtwomuupsi < 11$      & $80 < \mfourmu < 100$ \\
Dimuon vertex fit probability       & $>$5\%                                                &  $>$5\%  \\
Four-muon vertex fit probability    & $>$1\%                                                & \NA  \\
\end{scotch}
}
\vspace*{-0.15in}
}
{\renewcommand{\arraystretch}{1.2}
 \cmsTable{
\begin{scotch}{lcc}
Selection                           & \Zmumu + \Upsimumu                                    & \ZToFourMu     \\ \hline
All muons                           & $\abseta < 2.4$,\; $\StextIP<4$                       &  $\abseta < 2.4$,\; $\StextIP<4$     \\
Muon \pt [\GeVns{}]                 & $>$ 30, 15, 3, 3                                      & $>$ 20, 10, 5, 5   \\
Muon ID                             & \emph{tight}, \emph{tight}, \emph{soft}, \emph{soft}  & \emph{tight} for all muons \\
Isolation \isolation                & $<$ 0.35, 0.35, 0.7, 0.7                              & $<$0.35 for all muons \\
Mass ranges [\GeVns{}]              & $66 < \mtwomuPZ < 116$, $8.5 < \mtwomuupsi < 11$      & $80 < \mfourmu < 100$ \\
Dimuon vertex fit probability       & $>$5\%                                                &  $>$5\%  \\
Four-muon vertex fit probability    & $>$1\%    
\end{scotch}
}}
\label{tab:FiducialCutsTable}
\end{table*}

The ratio of fiducial cross sections $\RUpsi  = \frac{\sigmappZupsisBRs}{\sigmappZBRshort}$, evaluated in the fiducial phase space defined by the \pt and $\eta$ selections given in the first two rows of Table~\ref{tab:FiducialCutsTable}, where \sigmappZupsis is given by the sum of the two processes, SPS and DPS, and measured using the equation
\begin{linenomath}
\begin{equation} 
    \begin{aligned} 
        \RUpsi = \crosssectionformula.
    \end{aligned}
    \label{eq:XSecFormula}
\end{equation}
\end{linenomath}
The efficiencies $\epsilon$ are calculated using simulated event samples as the ratio of the number of events passing all of the analysis requirements applied to the reconstructed quantities, to the number of events passing the fiducial requirements applied to the generator-level quantities. 
They are $\epsilon_{\PZ}=48.5\%$, $\epsilon^\text{SPS}_{\Zupsis}=50.1\%$, and $\epsilon^\text{DPS}_{\Zupsis} = 42.4\%$. 
Theoretical uncertainties from the renormalization and factorization scales, and from imprecise knowledge of the parton distribution functions, largely cancel in the measured ratio. Using the results above we get \RUpsiresult.
We further extract \sigmaeff using the equation
\begin{linenomath}
\ifthenelse{\boolean{cms@external}}
{ 
\begin{multline}
\vspace*{-0.1in}
\sigmaeff = \frac{\sigma(\upsis)}{\RUpsiDPS} \frac{ \BZmumu \BUpsimumu} {\BZmumumumu} \\ 
\times \frac{\AZupsis}{\AZ},
\label{eq:two}
\vspace*{-0.1in}
\end{multline}
}
{
\begin{equation}
\sigmaeff = \frac{\sigma(\upsis)}{\RUpsiDPS} \frac{ \BZmumu \BUpsimumu} {\BZmumumumu} \frac{\AZupsis}{\AZ},
\label{eq:two}
\end{equation}
}
\end{linenomath}
where \RUpsiresultDPS is the DPS-only component of Eq.~\eqref{eq:XSecFormula}. The acceptances \AZupsis and \AZ are defined as the number of generated events passing the fiducial requirements, applied to the generator-level quantities, over the total number of generated events for each respective channel, and are $(37.6\pm 1.0)\%$ and $(26.3\pm0.8)\%$, respectively, where the uncertainties reflect the finite size of the simulated samples. Details about the event generation are provided in \suppMaterial. The \upsis meson fiducial production cross section calculated with \textsc{HELAC-Onia} to be $274 \pm 5\unit{nb}$~\cite{Lansberg:2016deg,Shao:2015vga,Shao:2012iz}. The values of \BZmumu and \BUpsimumu are taken from Ref.~\cite{PDG} and \BZmumumumu is from Ref~\cite{CMS:2026xzs}. The value of \sigmaeff is measured to be \sigmaeffresult. The determination of the asymmetric uncertainties and comparison with several other \sigmaeff measurements are discussed in \suppMaterial. 

Extending the inclusive measurement, \sigmaeff is also measured in three bins each of the \upsis meson and the \PZ boson \pt. Using the \textit{sPlot} weights obtained from the fit, we extract the signal \pt distributions of the \upsis meson and \PZ boson. The number of DPS events is obtained by subtracting the number of SPS events, which is estimated using the \pt distribution obtained from simulation scaled to match the total number of SPS events found in data \spsyield. Per-bin efficiencies are computed from simulation. 

\begin{figure}[tb!]
    \centering
    \includegraphics[width=\cmsFigSize]{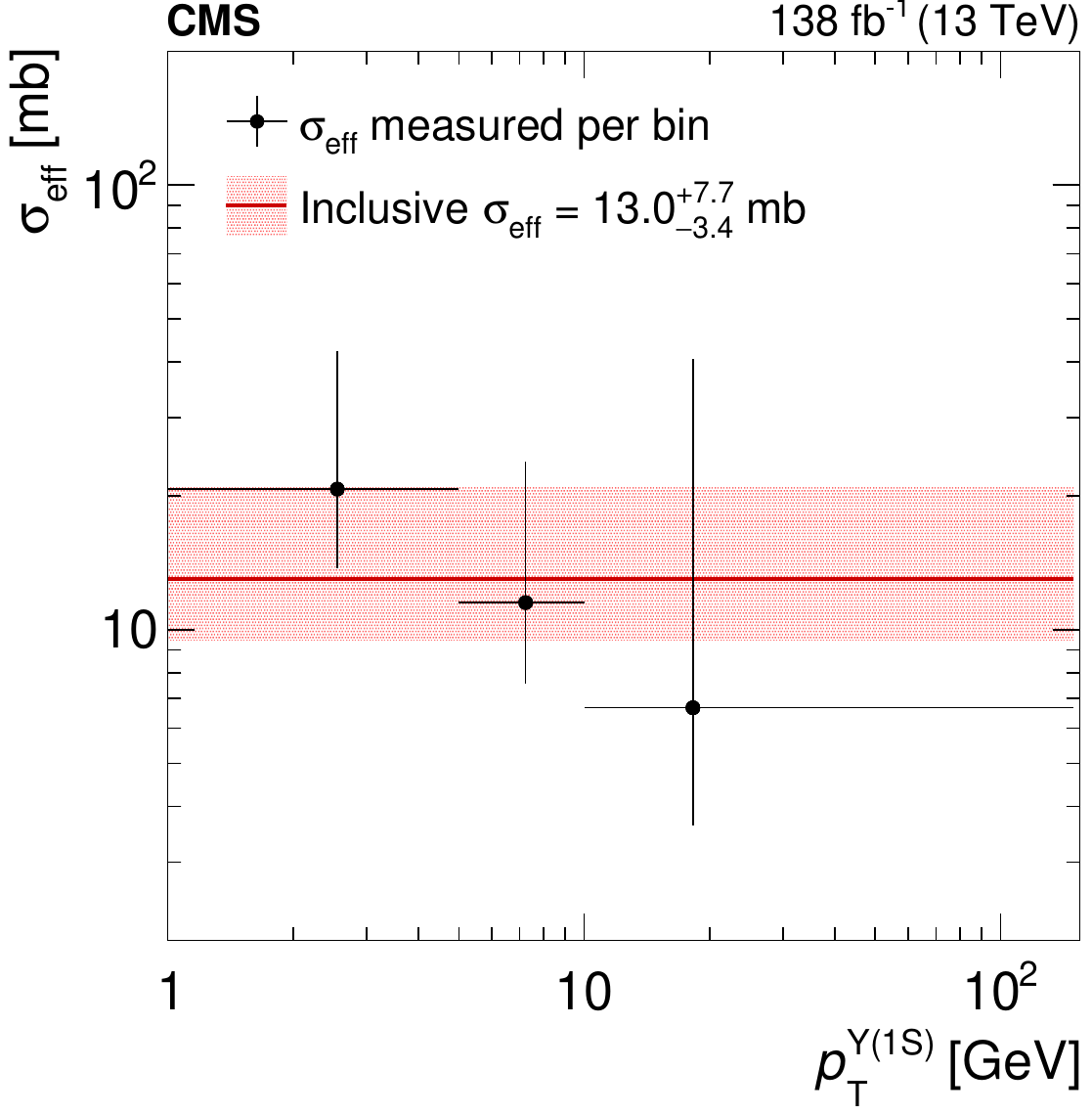}
    \includegraphics[width=\cmsFigSize]{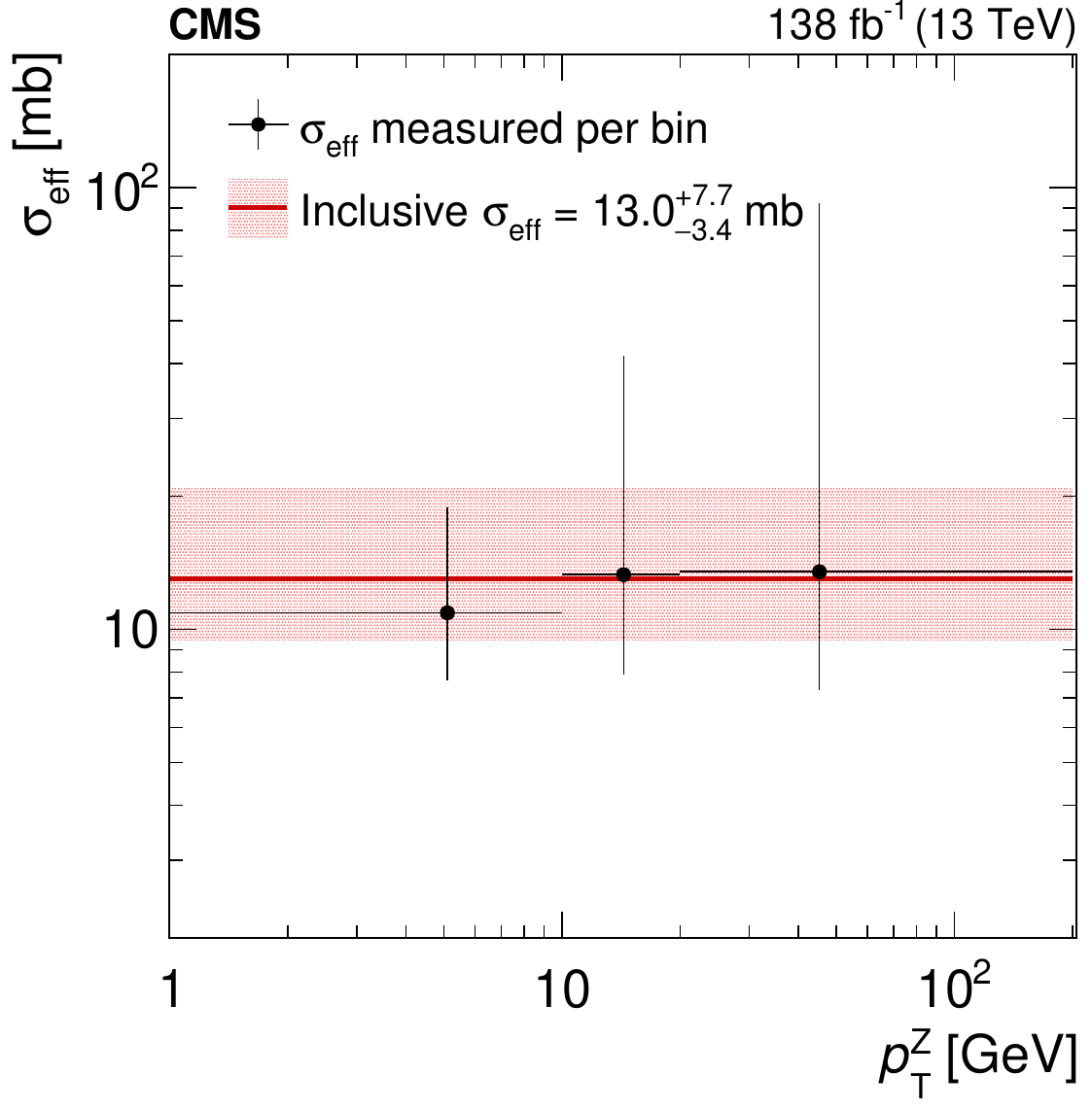} 
   \ifthenelse{\boolean{cms@external}}{\vspace*{-0.1in}}{}
 \caption{Measurements of \sigmaeff in bins of the \upsis meson (\cmsLeft) and \PZ boson (\cmsRight) \pt. The data are represented by the points with the black vertical bars showing the total uncertainties, which are dominated by the statistical uncertainty. The horizontal bars represent the bin widths, and the $x$ axis position of each marker corresponds to the weighted average of the \pt distribution within the respective bin. The red line indicates the nominal value found for the inclusive \sigmaeff, and the pink shaded region shows its total uncertainties.}
   \ifthenelse{\boolean{cms@external}}{\vspace*{-0.2in}}{}
    \label{fig:sigmaeffdiff}
\end{figure}

For the \upsis $\pt$-binned measurement, the full \ZToFourMu sample is used for each bin, and the \upsis meson production cross section in Eq.~\eqref{eq:two} is recalculated for each bin~\cite{Lansberg:2016deg,Shao:2015vga,Shao:2012iz} ($152 \pm 4$, $77 \pm 4$, and $45 \pm 4\unit{nb}$ for the 0--5, 5--10, and 10--150\GeV bins, respectively). For the \PZ \pt-binned measurement, the \ZToFourMu sample is divided into the same \pt bins, with corresponding \pt-dependent efficiencies in Eq.~\eqref{eq:XSecFormula}, while the \upsis meson production cross section in Eq.~\eqref{eq:two} is the same as for the inclusive measurement. The \sigmaeff measurements in bins of \upsis meson and \PZ boson \pt are shown in Fig.~\ref{fig:sigmaeffdiff}.

One might expect that, for the range of \pt probed, \sigmaeff would decrease as \upsis meson \pt increases and remain roughly independent of \PZ boson \pt. The former is because the higher-\pt \upsis mesons originate from gluons at higher Bjorken-$x$, which tend to be closer to the proton core, and, as a result, one observes a smaller \sigmaeff. In contrast, \sigmaeff as a function of the \PZ boson \pt could be expected to remain more or less constant, because in this case the scale is set by the mass of the \PZ boson, which is large in comparison to the changes in Bjorken-$x$ of the quarks responsible for the \PZ boson production.

In summary, the first study of the associated production of an \upsis meson and a \PZ boson in proton-proton collisions at $\sqrts = 13\TeV$ has been reported. Events with an \upsi meson and a \PZ boson, each decaying into a pair of opposite-sign muons, have been reconstructed. After all selection requirements, \yield events of \Zupsis production are observed. The statistical significance of the signal, relative to the background-only expectation, exceeds five standard deviations. The ratio of the fiducial cross section of $\Zupsis \to \mumumumu$ production to the fiducial cross section of \Zmumumumu production is measured to be \RUpsiresult. The derived DPS cross section ratio, \RUpsiresultDPS, can be expressed in terms of the effective DPS cross section $\sigmaeff = \sigmaeffresult$, which is closely related to the transverse distribution of partons in the proton. Such a value is consistent with \sigmaeff values obtained from double-quarkonium measurements in \pp and $\Pp\PAp$ collisions~\cite{CMS:2019jcb, ATLAS:2019jzd,Aaij:2016bqq, Abazov:2014qba,ATLAS:2014yjd,ATLAS:2014ofp,ATLAS:2014yjd,CDF:1993sbj}. For the first time, a binned measurement of the effective DPS cross section is performed, in bins of \upsis \pt or \PZ \pt. Such measurements provide extra information that can be used in studies of double-parton distribution functions~\cite{Manohar:2012jr}. 

\begin{acknowledgments}
We congratulate our colleagues in the CERN accelerator departments for the excellent performance of the LHC and thank the technical and administrative staffs at CERN and at other CMS institutes for their contributions to the success of the CMS effort. In addition, we gratefully acknowledge the computing centres and personnel of the Worldwide LHC Computing Grid and other centres for delivering so effectively the computing infrastructure essential to our analyses. Finally, we acknowledge the enduring support for the construction and operation of the LHC, the CMS detector, and the supporting computing infrastructure provided by the following funding agencies: SC (Armenia), BMBWF and FWF (Austria); FNRS and FWO (Belgium); CNPq, CAPES, FAPERJ, FAPERGS, and FAPESP (Brazil); MES and BNSF (Bulgaria); CERN; CAS, MoST, and NSFC (China); MINCIENCIAS (Colombia); MSES and CSF (Croatia); RIF (Cyprus); SENESCYT (Ecuador); ERC PRG, TARISTU24-TK10 and MoER TK202 (Estonia); Academy of Finland, MEC, and HIP (Finland); CEA and CNRS/IN2P3 (France); SRNSF (Georgia); BMFTR, DFG, and HGF (Germany); GSRI (Greece); NKFIH (Hungary); DAE and DST (India); IPM (Iran); SFI (Ireland); INFN (Italy); MSIT and NRF (Republic of Korea); MES (Latvia); LMTLT (Lithuania); MOE and UM (Malaysia); BUAP, CINVESTAV, CONACYT, LNS, SEP, and UASLP-FAI (Mexico); MOS (Montenegro); MBIE (New Zealand); PAEC (Pakistan); MES, NSC, and NAWA (Poland); FCT (Portugal); MESTD (Serbia); MICIU/AEI and PCTI (Spain); MOSTR (Sri Lanka); Swiss Funding Agencies (Switzerland); MST (Taipei); MHESI (Thailand); TUBITAK and TENMAK (T\"{u}rkiye); NASU (Ukraine); STFC (United Kingdom); DOE and NSF (USA).    

\end{acknowledgments}\section*{Data availability} Release and preservation of data used by the CMS Collaboration as the basis for publications is guided by the  \href{https://doi.org/10.7483/OPENDATA.CMS.1BNU.8V1W}{CMS data preservation, re-use and open access policy}.

\bibliography{auto_generated}

\appendix
\ifthenelse{\boolean{cms@external}}{\section{}}{\section{End Matter}}
\label{app:suppMat}
The CMS apparatus~\cite{CMS:2008xjf,CMS:2023gfb} is a multipurpose, nearly hermetic detector, designed to trigger on~\cite{CMS:2020cmk,CMS:2016ngn} and identify electrons, muons, photons, and (charged and neutral) hadrons~\cite{CMS:2020uim,CMS:2018rym,CMS:2014pgm}. A global ``particle-flow'' algorithm~\cite{CMS:2017yfk} aims to reconstruct all the individual particles in an event by combining information provided by the all-silicon inner tracker and by the crystal electromagnetic, and the brass and scintillator hadron calorimeters, operating inside a 3.8\unit{T} superconducting solenoid, with data from the gas-ionization muon detectors embedded in the flux-return yoke outside the solenoid. Muons are measured in a range $\abseta < 2.4$, with detection planes made using three technologies: drift tubes, cathode strip chambers, and resistive-plate chambers. The efficiency to reconstruct and identify muons is greater than 96\%. Matching muons to tracks measured in the silicon tracker results in a relative \pt resolution for muons with \pt up to 100\GeV, of 1\% in the barrel and 3\% in the endcaps.

Events of interest are selected using a two-tiered trigger system. The first level, composed of custom hardware processors, uses information from the calorimeters and muon detectors to select events at a rate of around 100\unit{kHz} within a fixed latency of 4\mus~\cite{CMS:2020cmk}. The second level, known as the high-level trigger, consists of a farm of processors running a version of the full event reconstruction software optimized for fast processing, and reduces the event rate to a few kHz before data storage~\cite{CMS:2016ngn,CMS:2024aqx}. The data were collected with a suite of double- and triple-muon triggers. Thresholds for the double-muon (triple-muon) triggers are 17 and 8\GeV (12, 10, and 5\GeV) for all three data-taking years. In 2017--2018, an additional triple-muon trigger with thresholds of 10, 5, and 5\GeV was also employed. The triggers are nearly 100\% efficient for events passing the offline selection.

\subsection*{Simulated samples}\label{app:SimSamples}
For the signal process, two sets of Monte Carlo simulated event samples are produced: a DPS set, simulated at leading order~(LO) using \PYTHIA 8.240~\cite{Sjostrand:2014zea}, and an SPS set, created using the next-to-LO~(NLO) \textsc{HELAC-Onia} 2.5.5~\cite{Shao:2015vga,Shao:2012iz} parton-level event generator, interfaced with \PYTHIA for the description of fragmentation and hadronization. Simulated DPS events contain \upsis mesons produced promptly, as well as events in which they come from ``feed-down'', or the decays of more massive states, namely \UpsiTwoS and \UpsiThreeS mesons, as well as \PGcbJ{1P} mesons, with total angular momenta $J=0$, 1, and 2. The SPS event generation is made under the assumption that the \upsis mesons produced in association with a \PZ boson, in the $\Pp\Pp \to \upsis \PZ$ process, are dominantly produced from color singlet $\PQb\PAQb$ pairs, so that no color-octet terms are generated due to their expected smallness. Furthermore, only direct \upsis meson production is considered, after having determined from simulation studies that accounting for feed-down decays of heavier bottomonium states leads to a negligible effect on the variables of interest, the rapidity difference \deltay and the azimuthal opening angle \deltaphi between the \upsis meson's and the \PZ boson's momentum directions.

For the normalization channel, the matrix elements for the \ZToFourMu events are computed at NLO using \POWHEG 2.0~\cite{Nason:2004rx,Frixione:2007vw,Alioli:2010xd,Melia:2011tj,Nason:2013ydw}, with the parton showering and hadronization described by \PYTHIA. 

For all samples, the CP5 underlying event tune~\cite{CMS:2022awf} and the next-to-NLO NNPDF3.1~\cite{NNPDF:2017mvq} parton distribution functions are used, while the detector response is simulated with the \GEANTfour~\cite{Agostinelli:2002hh} package. The effect of pileup is included by overlaying simulated minimum bias events on the hard-scattering process, with the multiplicity profile matching that observed in data. 

\ifthenelse{\boolean{cms@external}}{  
\begin{figure*}[htb!!]
\centering
\includegraphics[width=0.67\textwidth]{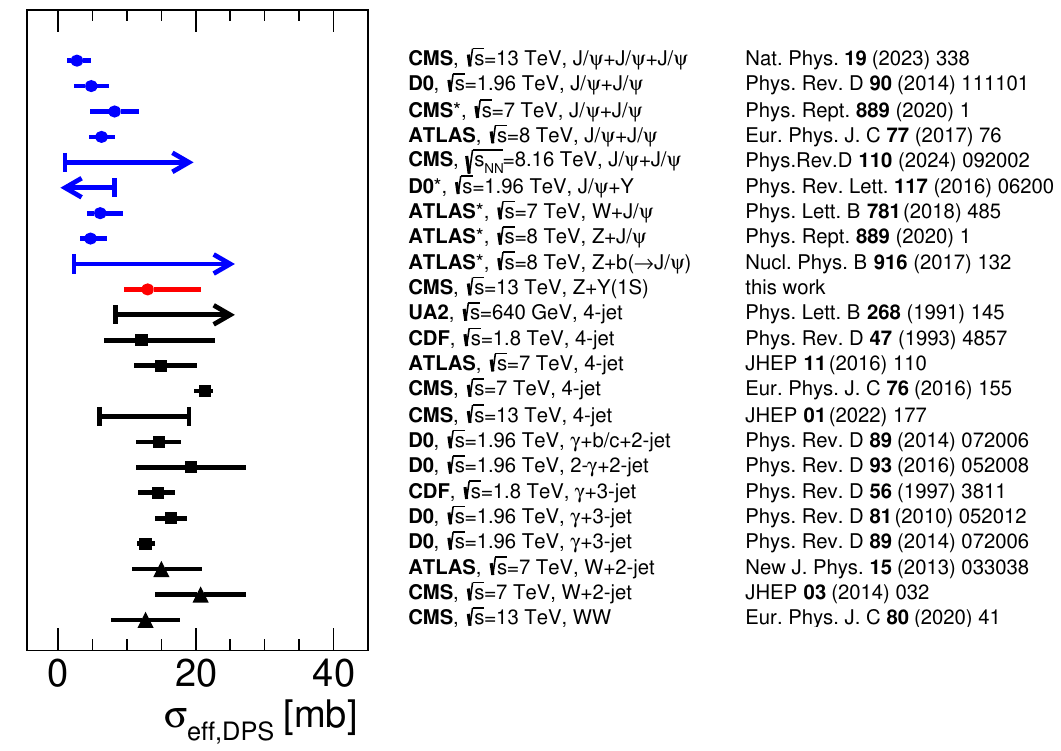}
\caption{A summary of \sigmaeff results measured using quarkonia (blue) and harder objects (black). The present result is shown in red. Arrows indicate that only the lower limits on \sigmaeff are available. The error bar without the central value corresponds to a range of \sigmaeff extractions using different MC models from a CMS 4-jet analysis~\cite{CMS:2021lxi}. For the results marked with an asterisk, more recent theoretical reinterpretations based on improved calculations of the corresponding SPS cross section are plotted.}
\label{fig:SigmaEffComp}
\end{figure*}
}{}

\subsection*{Systematic uncertainties}\label{app:SystUncerts}
The uncertainty from the limited number of data events is dominant for the measurements presented in this Letter. Nevertheless, we carefully evaluate various systematic uncertainties and summarize them in this section.

We examine the choice of the \upsins signal model by changing the lineshape description from a single Gaussian function for each of the three \upsi states to a double-Gaussian function. We observe a difference in the \Zupsis signal yield of \UpsiSigModUnc. We next explore the choice of the parameterization of the \upsins background by replacing the default exponential function with a first-order polynomial function. The difference in the \Zupsis yield between the two methods is \UpsiBkgModUnc.

We change the \PZ boson signal model from a Voigtian function to a sum of Voigtian and Gaussian functions, where the additional Gaussian function is meant to account for the fact that the detector resolution is not constant but varies with the kinematics of the muons. In the signal channel, this results in a \ZSigModUncZUpsi change in the \Zupsis yield, while in the \ZToFourMu channel, the corresponding change in the event yield is \ZSigModUncZToMuMuMuMu. Changing the \PZ candidate background model from an exponential function to a first-order polynomial function leads to a difference of \ZBkgModUncZUpsi in the \Zupsis yield, while it results in a difference of \ZBkgModUncZToMuMuMuMu in the \ZToFourMu yield.

The overall effect of the limited event counts in simulated samples on the \RUpsi calculation, where \RUpsi is defined by Eq.~\eqref{eq:XSecFormula}, is \MCStatsUncRatio. The effect is also \MCStatsUncSigmaEff on the inclusive \sigmaeff computation.

There are also uncertainties that arise from the limited knowledge of the various branching fractions used in the calculations. The uncertainties in the world-average \Upsimumu and \Zmumu branching fractions are \BFUpsiMuMuUnc and \BFZMuMuUnc, respectively~\cite{PDG}. The uncertainty in the \ZToFourMu branching fraction, per Ref.~\cite{CMS:2026xzs}, which is a detailed CMS study of $\PZ \to 4 \Pell$ events, is \BFZMuMuMuMuUnc. 

Uncertainties associated with the trigger, muon reconstruction and identification, and vertexing efficiency cancel in the ratio within the precision of the measurement.

The final source of systematic uncertainty relates to the \upsis production cross section calculation. The \upsis production cross section, $274 \pm 5\unit{nb}$~\cite{Lansberg:2016deg,Shao:2015vga,Shao:2012iz}, is taken from the \textsc{HELAC-Onia} event generator and is used in the inclusive \sigmaeff calculation. On top of this, we take the maximum difference between the theoretical fits to data at 7\TeV~\cite{Lansberg:2016deg,Shao:2015vga} and the CMS measurement of this cross section at the same energy~\cite{CMS:2013qur}, and assign this relative difference as an additional uncertainty in the \upsis production cross section at 13\TeV. This uncertainty applies to both the inclusive and binned \sigmaeff results and is taken to be \UpsiXSecTheoryUnc.

The total systematic uncertainty for both the \RUpsi and the \sigmaeff measurements is given by the sum in quadrature of the systematic uncertainties described in this section. The overall systematic uncertainty is \RatioTotalSystUncert for the \RUpsi result and \SigmaEffTotalSystUncert for the \sigmaeff result.

For the \pt-binned measurements, uncertainties pertaining to the lineshapes chosen to describe the signal and background distributions are evaluated at the per-bin level; the alternative shapes chosen are the same as for the inclusive measurement. The uncertainties due to the finite sizes of the simulated samples are also calculated and applied bin by bin, while the uncertainties in the branching fractions used in the computations are common for all bins. Finally, in the case of the measurements binned in \upsis \pt, uncertainties related to the use of the \upsis cross sections are taken from \textsc{HELAC-Onia} at a per-bin level.

\ifthenelse{\boolean{cms@external}}{}{ 
\begin{figure*}[htb!]
\centering
\includegraphics[width=\textwidth]{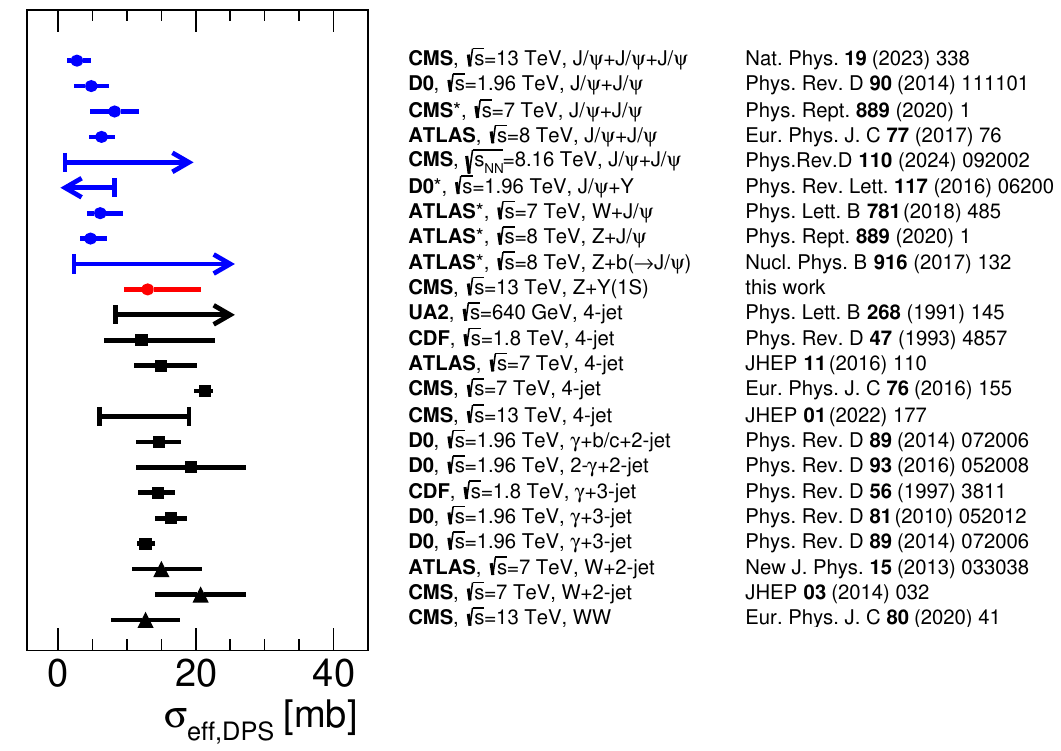}
\caption{A summary of \sigmaeff results measured using quarkonia (blue) and harder objects (black). The present result is shown in red. Arrows indicate that only the lower limits on \sigmaeff are available. The error bar without the central value corresponds to a range of \sigmaeff extractions using different MC models from a CMS 4-jet analysis~\cite{CMS:2021lxi}. For the results marked with an asterisk, more recent theoretical reinterpretations based on improved calculations of the corresponding SPS cross section are plotted.}
\label{fig:SigmaEffComp}
\end{figure*}}

\subsection*{Summary of \texorpdfstring{\boldmath{\sigmaeff}}{sigma-eff} measurements}\label{app:SummaryOfSigmaEffs}
The inclusive \sigmaeff result is compared with several previous \sigmaeff measurements in Fig.~\ref{fig:SigmaEffComp}. We note that, while the ratios ${\mathcal R}$ have symmetric uncertainties because they depend linearly on the event yields in the signal channel, \sigmaeff values depend on the inverse of a typically small corresponding signal yield, leading to asymmetric uncertainties.
To calculate the asymmetric uncertainties in the \sigmaeff measurements, we start from the values of the measurements' inverses,  which have symmetric uncertainties. We next add or subtract each symmetric uncertainty from each central value, and then take the inverse of each of these results to get the lower and upper bounds, respectively, leading to asymmetric uncertainties.

\cleardoublepage \section{The CMS Collaboration \label{app:collab}}\begin{sloppypar}\hyphenpenalty=5000\widowpenalty=500\clubpenalty=5000\cmsinstitute{Yerevan Physics Institute, Yerevan, Armenia}
{\tolerance=6000
A.~Hayrapetyan, V.~Makarenko\cmsorcid{0000-0002-8406-8605}, A.~Tumasyan\cmsAuthorMark{1}\cmsorcid{0009-0000-0684-6742}
\par}
\cmsinstitute{Institut f\"{u}r Hochenergiephysik, Vienna, Austria}
{\tolerance=6000
W.~Adam\cmsorcid{0000-0001-9099-4341}, J.W.~Andrejkovic, L.~Benato\cmsorcid{0000-0001-5135-7489}, T.~Bergauer\cmsorcid{0000-0002-5786-0293}, M.~Dragicevic\cmsorcid{0000-0003-1967-6783}, C.~Giordano, P.S.~Hussain\cmsorcid{0000-0002-4825-5278}, M.~Jeitler\cmsAuthorMark{2}\cmsorcid{0000-0002-5141-9560}, N.~Krammer\cmsorcid{0000-0002-0548-0985}, A.~Li\cmsorcid{0000-0002-4547-116X}, D.~Liko\cmsorcid{0000-0002-3380-473X}, M.~Matthewman, I.~Mikulec\cmsorcid{0000-0003-0385-2746}, J.~Schieck\cmsAuthorMark{2}\cmsorcid{0000-0002-1058-8093}, R.~Sch\"{o}fbeck\cmsAuthorMark{2}\cmsorcid{0000-0002-2332-8784}, D.~Schwarz\cmsorcid{0000-0002-3821-7331}, M.~Shooshtari\cmsorcid{0009-0004-8882-4887}, M.~Sonawane\cmsorcid{0000-0003-0510-7010}, W.~Waltenberger\cmsorcid{0000-0002-6215-7228}, C.-E.~Wulz\cmsAuthorMark{2}\cmsorcid{0000-0001-9226-5812}
% ...
\par}
\vskip\cmsinstskip
\dag:~Deceased\\
$^{1}$Also at Yerevan State University, Yerevan, Armenia\\
$^{2}$Also at TU Wien, Vienna, Austria\\
% ...
\end{sloppypar}
\end{document}